\documentclass[aps,pre,twocolumn,superscriptaddress]{revtex4-2}\pdfoutput=1
\usepackage{amsmath,amssymb}
\usepackage{graphics,graphicx,color}
\usepackage{dcolumn,bm}
\usepackage{tikz}
\usepackage{natbib}

\usepackage[utf8]{inputenc}
\usepackage[colorlinks=true,citecolor=blue,urlcolor=blue]{hyperref}
\def\[{\left[}
\def\]{\right]}
\def\({\left(}
\def\){\right)}
\def\be{\begin{equation}}
\def\ee{\end{equation}}
\def\bea{\begin{eqnarray}}
\def\eea{\end{eqnarray}}

\newcommand{\gaug}
{\affiliation{Institute for Theoretical Physics, Georg-August-University G\"ottingen, 37077 G\"ottingen, Germany}}

\begin{document}
\title{How to Study a Persistent Active Glassy System}

\author{Rituparno Mandal}%
\email[Email: ]{rituparno.mandal@uni-goettingen.de}
\gaug

\author{Peter Sollich}%
\email[Email: ]{peter.sollich@uni-goettingen.de}
\gaug
\affiliation{Department of Mathematics, King's College London, Strand, London WC2R 2LS, UK}

\begin{abstract}
We explore glassy dynamics of dense assemblies of soft particles that are self-propelled by active forces. These forces have a fixed amplitude and a propulsion direction that varies on a timescale $\tau_p$, the persistence timescale. Numerical simulations of such active glasses are computationally challenging when the dynamics is governed by large persistence times. We describe in detail a recently proposed scheme that allows one to study directly the dynamics in the large persistence time limit, on timescales around and well above the persistence time.  We discuss the idea behind the proposed scheme, which we call ``activity-driven dynamics'', as well as its numerical implementation. We establish that our prescription faithfully reproduces all dynamical quantities in the appropriate limit $\tau_p\to\infty$. We deploy the approach to explore in detail the statistics of Eshelby-like plastic events in the steady state dynamics of a dense and intermittent active glass.
\end{abstract}

\maketitle

\section{Introduction}

Disordered or amorphous solids, also known as glasses, are one of the most abundant states of matter~\cite{anderson95}, but remain less well understood than their closest relatives, {\it i.e.} liquids and crystals~\cite{berthierrmp11}. Many theoretical approaches has been put forward over the last few decades, including Mode Coupling Theory~\cite{spdas04}, Random First Order Transition Theory{~\cite{thirumalai15}}, Free Volume Theory~\cite{dyre06}, and more recently exact solutions in infinite dimensions for hard sphere glasses~\cite{parisi17}. Alongside these efforts, many significant and fundamental discoveries have been made in  the numerical investigation of model glass formers~\cite{berthierrmp11}. Still, in spite of the enormous amount of research done in recent decades, a complete understanding of this disordered solid phase remains elusive~\cite{berthierrmp11,berthier20}.

Compared to the physics of glasses, {\em active matter} is a relatively recent field of study that lies at the intersection of soft matter, non-equilibrium statistical mechanics and biological systems~\cite{marchetti13,volpe16}.
This field has emerged as one of the most fruitful areas of research in the last decade. Being inherently out of equilibrium~\cite{sriram10}, active matter systems show fascinating dynamical phases (swirls or vortices) and ordering (swarms, flocks, active nematic states etc.), giant number fluctuations and intriguing mechanical and dynamical responses~\cite{marchetti13,volpe16}.

Active matter systems can exhibit gas, liquid, liquid crystal and crystalline phases~\cite{marchetti13,volpe16} but also {\em active glasses}. These are a dense and disordered form of active matter, and our focus in this paper. They sit at the intersection of the fields of glass physics and active matter, and are relevant to understanding synthetic active materials such as dense assemblies of Janus colloids~\cite{leomach19,leomach19b} as well as 
many biological systems including the cytoplasm~\cite{parry14}, with its ATP dependent molecular activity, or epithelial tissues~\cite{angelini11}, which are dense collections of motile cells.

\begin{figure}
\includegraphics[height = 0.85\linewidth]{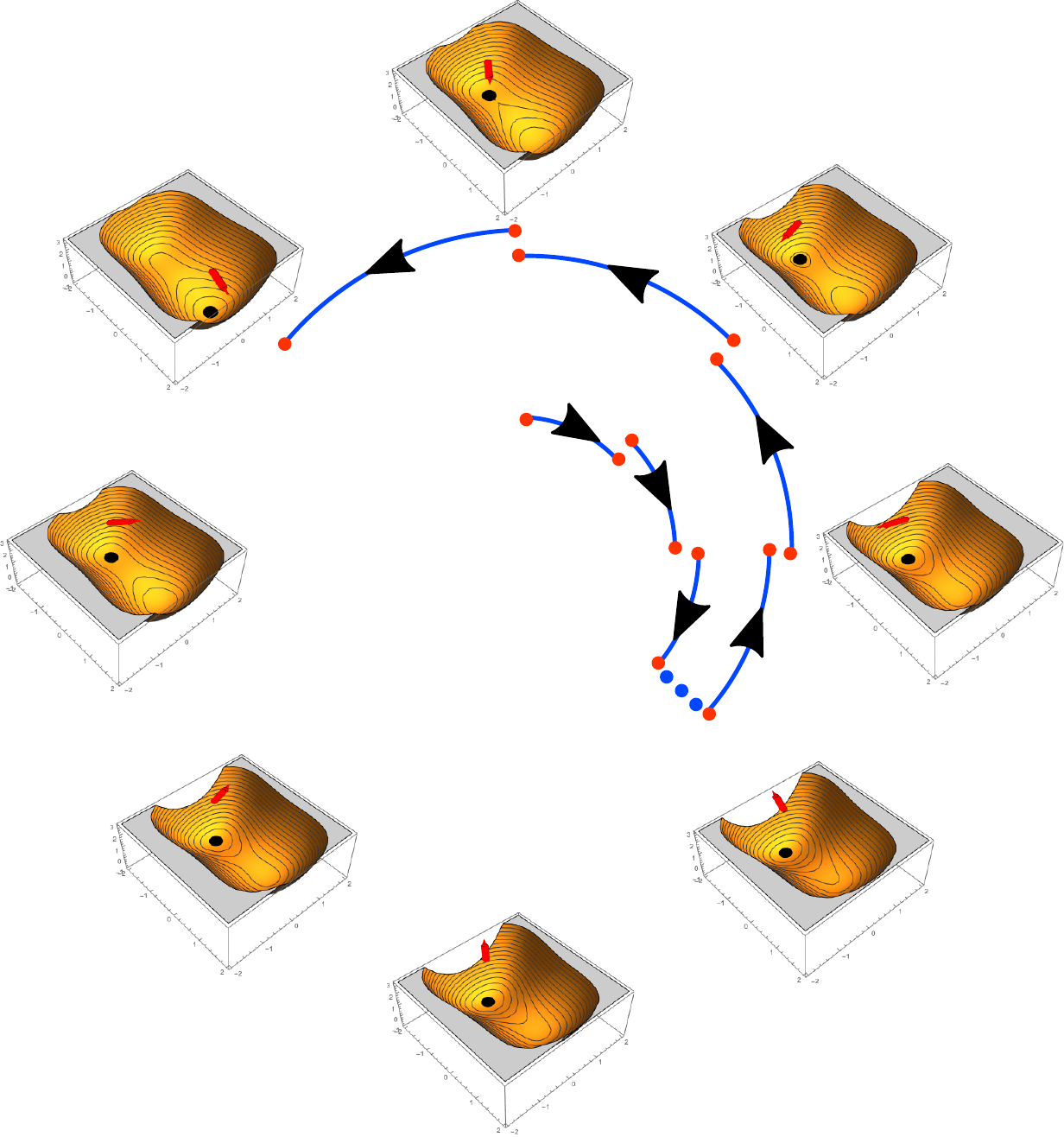}
\caption{Schematic of particle motion resulting from the random tilting of the effective potential landscape by active forces, in a cartoon showing a two-dimensional landscape $V^{\rm eff}(x_1,x_2)$. The blue arrows in the centre indicate the time evolution (inside to out) of a propulsion force direction in small steps of rescaled time $t'=t/\tau_p$. Each change in this direction (indicated by red arrows above the landscape) slightly changes the tilt of the effective potential and thus the particle configuration (black dot), which is always at a local minimum of $V^{\rm eff}$. In the last step shown, the change in tilt destabilizes the current local minimum and a plastic event takes place, with a significant rearrangement of the particle configuration into a new local minimum.}
\label{fig:tilting}
\end{figure}

Recent studies on active glasses (see Ref.~\cite{janssen19} for a comprehensive review) have revealed many interesting phenomena. Glass transition boundaries have been found to shift with increasing activity, for example, towards higher area fractions~\cite{silke11,ni13,berthier14} or lower temperature in density or temperature-controlled glasses~\cite{berthier13,mandal16}, respectively. Similarities but also substantial differences to passive glasses have been reported, including a novel intermittent dynamical phase~\cite{mandal20} and two-step aging scenarios~\cite{mandalprl20}. Theoretical progress in  understanding this actively driven solid state of matter has also been made, with the development of Mode Coupling Theory for dense active systems~\cite{nandi17}, active Random First Order Transition theory~\cite{nandi18} and active trap models~\cite{nir20}, to name a few.

A general observation from existing studies is that in the limit of weak (more precisely, weakly persistent) activity, active glasses behave essentially like passive thermal systems with an effective temperature~\cite{berthier13,mandal16,mandal20,mandalprl20}. Strong departures from thermal behaviour appear in the opposite limit of highly persistent activity. Our aim in this article is to set out in detail a recently proposed method~\cite{mandalprl20} that allows for the efficient simulation of such ``extreme active matter''~\cite{mandal20}. We refer to this approach as Activity Driven Dynamics (ADD). By comparing a range of  dynamical quantities we establish that the new algorithm can capture reliably the asymptotic behaviour for $\tau_p \to \infty$ while remaining computationally efficient. Finally we deploy the method to study the statistics of Eshelby-like plastic rearrangements seen in the steady state dynamics of an active glass.

\begin{figure}
\includegraphics[height = 1.3\linewidth]{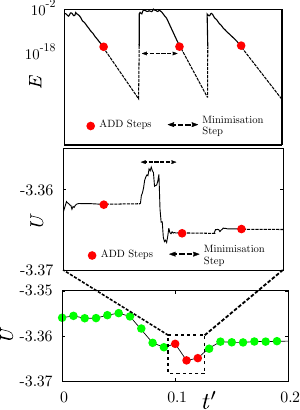}
\caption{ (Top) A typical time series of kinetic energy (per particle) in the ADD algorithm where red points represents the state of the system after each ADD step and the black segments (whose duration is indicated by the double dashed arrows) show the minimisation steps, which are of varying length due to their adaptive in nature. For large $\tau_p$, the length of the minimisation step becomes negligible compared to the duration of the ADD steps. (Middle) Corresponding values of potential energy per particle, $U=V/N$; note that the linear terms from the active forces are not included so that $U$ does not decrease monotonically during minimization. (Bottom) The resulting potential energy time series in scaled time $t'=t/\tau_p$, showing in red the results from the three successive ADD steps in the top and middle panels.}
\label{fig:ke}
\end{figure}

\section{Activity driven dynamics}

We consider in the following systems of active particles moving in $d=2$ dimensions with a propulsion force that remains fixed in magnitude but changes randomly in time~\cite{marchetti12,takatori15,levis17,solon18}. Assuming inertial dynamics with friction against a stationary solvent then gives the equations of motion
\begin{equation}
m{\ddot{\mathbf{r}}}_i=-\gamma \dot{\bf{r}}_i + \mathbf{f}_{i} + f \mathbf{n}_i  
\label{eqom2}
\end{equation}
Here ${\bf{r}}_i$ is the position vector of particle $i$ ($i=1,\ldots,N$), $m$ is the particle mass, $\gamma$ is the friction coefficient and $\mathbf{f}_{i}=-\nabla_i V$ is the total interaction force on particle $i$ derived from some potential $V$. We will use a sum of pairwise Lennard-Jones interactions below but $V$ can in general contain arbitrary many-body interactions.

A key parameter for the physical behaviour is $f$, which measures the strength -- assumed constant in time -- of the propulsion force on each particle. The direction of this propulsion force, $\mathbf{n}_i$, is a unit vector
\begin{equation}
\mathbf{n}_i \equiv (\cos{\theta_i},\sin{\theta_i})
\end{equation}
which is assumed to perform rotational Brownian motion with timescale $\tau_p$:
\begin{equation}
   \dot{ \theta_i}=\sqrt{{2}/{\tau_p}}\, \eta_i
    \label{eqom3}
\end{equation}
Here $\eta_i$ is zero mean Gaussian white noise with correlator $\langle\eta(t_1)\eta(t_2)\rangle=\delta(t_1-t_2)$.

Our focus in the following will be the limit of large $\tau_p$, {\em i.e.}\ of a highly persistent active glass. Such a glass can arise if the system is dense enough, and the active propulsion force $f$ not too large. If under these conditions we fix the directions ${\bf n}_i$ of the self-propulsion forces, then the  time evolution of the particle positions, Eq.(\ref{eqom2}), will rapidly reach an arrested state where the total force on each particle vanishes. 
Now if $\tau_p$ is large but finite, the propulsion force orientations ${\bf n}_i$ will change on a timescale of $\tau_p$. On the other hand,
the time for the particles to reach an arrested state for any given set of ${\bf n}_i$ does {\em not} grow with $\tau_p$. In the limit $\tau_p \to \infty$, the particle configuration thus tracks the propulsion forces effectively instantaneously and we have Activity Driven Dynamics (ADD): the time evolution of the system is driven only by changes in the active forces. The reason why this limiting dynamics is useful for numerical simulation is that for each time step of $O(\tau_p)$ that corresponds to a small change of the $\{{\bf n}_i\}$, we only need to simulate for a time that does not scale with $\tau_p$, until the particles have reached their arrested state given the new $\{{\bf n}_i\}$. Thus in the limit large $\tau_p$ we expect a reduction in computational effort by a factor of order $\tau_p$, if we work in units were typical relaxation times are of order unity.

\begin{figure}
\includegraphics[height = 0.7\linewidth]{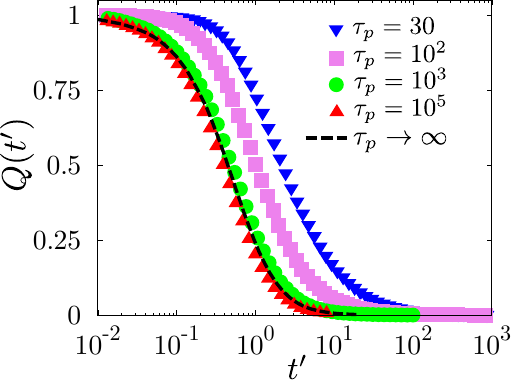}
\caption{Two-point overlap correlation function $Q(t)$ as a function of scaled time $t'=t/\tau_p$, calculated for active force amplitude $f=1.2$ and for different $\tau_p$ as shown. The results converge for large $\tau_p$ and approach the prediction of ADD (black dashed line), confirming the validity of the method.}
\label{fig:qt}
\end{figure}

To derive ADD more formally, we rescale time to $t'=t/\tau_p$ and write the equations of motion Eq.(\ref{eqom2},\ref{eqom3}) in this new time variable:
\begin{eqnarray}
m \frac{1}{\tau_p^2}\,\frac{d^2 \mathbf{r}_i}{d{t^{\prime}}^2}&=&-\gamma\frac{1}{\tau_p}\, \frac{d \mathbf{r}_i}{dt^{\prime}} +  \mathbf{f}_{i} + f \mathbf{n}_i
\label{scaled2}\\
\frac{d \theta_i}{dt^{\prime}}&=&\sqrt{2} \,\eta'_i.
 \label{scaled3}
\end{eqnarray}
Here $\eta_i'$ is a scaled white noise defined to have unit variance in the scaled time variables, {\em i.e.}\ $\langle \eta_i'(t'_1)\eta_i'(t'_2)\rangle = \delta(t'_1-t'_2)$, which gives $\eta'_i = \sqrt{\tau_p}\eta_i$. Now the basic assumption of ADD is that for large $\tau_p$, the particle dynamics is driven by that of the ${\bf n}_i$ (or equivalently $\theta_i$), so that the evolution of the particle configuration takes place on timescales of $O(\tau_p)$. The derivatives w.r.t.\  the rescaled time $t'$, $d{\bf r}_i/dt'$ and $d{\bf r}_i^2/dt'{}^2$, must then remain {\em finite} as $\tau_p\to\infty$. Eq.(\ref{scaled2}) thus implies that for $\tau_p\to\infty$ 
\begin{equation}
0=  \mathbf{f}_{i} + f \mathbf{n}_i.  
\label{minimize}
\end{equation}
Together with Eq.(\ref{scaled3}) this equation defines ADD: in rescaled time $t'$, the Brownian dynamics of the propulsion force orientations has a fixed rotational diffusion constant independently of $\tau_p$, while the particle configuration simply tracks the evolution of the $\{{\bf n}_i\}$ so that the total force on every particle vanishes at all times (cf.\ Eq.(\ref{minimize})). The latter condition can also be phrased as saying that the particle configuration always locally minimizes an effective potential that has been {\em tilted} by the active forces,
\begin{equation}
    V^{\rm eff} = V-\sum_i f{\bf n}_i \cdot {\bf r}_i
\label{Veff_def}
\end{equation}
As the active force directions ${\bf n}_i$ evolve, so does $V^{\rm eff}$. In a small step $\delta t'$ of rescaled time $t'$, the system can then either remain in a smoothly evolving minimum of $V^{\rm eff}$, or the existing minimum can become unstable and we will observe a {\em plastic event} where particles rearrange irreversibly and effectively instantaneously when measured in rescaled time $t'$. Fig.~\ref{fig:tilting} illustrates the distinction between these two types of motion with a sketch for a two-dimensional particle configuration space.

Before discussing specific models and the computational implementation of ADD, we comment briefly on the generality of the approach. Clearly the reasoning behind ADD can be applied equally well in $d=3$ dimensions, where the propulsion force directions ${\bf n}_i$ then perform a Brownian walk on a unit sphere rather than a circle as in $d=2$. Other models of active propulsion can also be treated for $\tau_p\to\infty$, including Active Ornstein-Uhlenbeck particles. Their equation of motion in the overdamped case can be written as~\cite{maggi14,marconi15}
\begin{eqnarray}
    \gamma\dot{\mathbf{r}}_i&=& \mathbf{f}_i+\gamma\mathbf{v}_i
\\    \tau_p \dot{\mathbf{v}}_i&=&-\mathbf{v}_i+\sqrt{2D}\, \boldsymbol{\eta}_i
\end{eqnarray}
where ${\bf v}_i$ is the active velocity and $D$ the translational diffusion constant. The noise $\boldsymbol{\eta}_i$ is now vectorial, with zero mean and covariance  $\langle \eta_{i \alpha}(t_1) \eta_{j \beta}(t_2) \rangle=\delta_{ij} \delta_{\alpha \beta} \delta(t_1-t_2)$, with $\alpha,\beta$ labelling the Cartesian components.
In the above way of writing the dynamics, the steady state variance of each component of the active force $\gamma{\bf v}_i$ is given by $\tilde f^2=\gamma^2 D/\tau_p$. To identify this scale explicitly we write $\gamma{\bf v}_i=\tilde{f}\tilde{\bf n}_i$ so that $\tilde{\bf n}_i$ will be a vector with length of order unity.  
Scaling time by $\tau_p$ as we did previously yields
\begin{eqnarray}
   \gamma \frac{1}{\tau_p} \frac{d \mathbf{r}_i}{d t^{\prime}}&=& \mathbf{f}_i+\tilde f\tilde{\mathbf{n}}_i\\
 \frac{d \tilde{\mathbf{n}}_i}{d t^{\prime}}&=&- \tilde{\mathbf{n}}_i+\sqrt{2}\,{\boldsymbol{\eta}}^{\prime}_i
 \end{eqnarray}
with $\boldsymbol{\eta}_i'=\sqrt{\tau_p}\,\boldsymbol{\eta}_i$ the corresponding rescaled noise. The ADD limit $\tau_p\to\infty$ (at constant $\tilde f$) now gives again a limiting dynamics for the active force directions $\tilde{\bf n}_i$, with the force-free condition $0={\bf f}_i+\tilde f\tilde{\bf n}_i$ determining the evolution of the particle configuration.

\section{Model for Simulation}

For numerical modelling we use in this paper the widely studied Kob-Andersen model glass former~\cite{kob95,bruning08}. All our simulations are performed in $d=2$ dimensions with a number of particles between $N=1000$ and $4000$ in a square periodic box. The self-propulsion force on each particle has fixed magnitude $f$ and diffusive orientational dynamics~\cite{mandal20} as discussed above. The net interaction force on particle $i$ is $\mathbf{f}_{i}=\sum_{j} \mathbf{f}_{ij}$ where $\mathbf{f}_{ij}$ is a pairwise interaction force derived from a Lennard-Jones potential:
\begin{equation}
 V_{ij}(r)=4 \epsilon_{\alpha \beta} \left[\left(\frac{\sigma_{\alpha \beta}}{r}\right)^{12}-\left(\frac{\sigma_{\alpha \beta}}{r}\right)^{6}\right]
\end{equation} 
where $r=|\mathbf{r}_i-\mathbf{r}_j|$ is the distance between particles $i$ and $j$. The model contains a mixture of $A$- and $B$-type particles and the interaction parameters depend on the type ($\alpha,\beta$) of particles involved. The number ratio ($A:B$) of particles is $65:35$ and we have chosen the values of $\sigma_{\alpha
\beta}$ and $\epsilon_{\alpha \beta}$ to be: $\sigma_{AB}=0.8
\sigma_{AA}$, $\sigma_{BB}=0.88 \sigma_{AA}$, $\epsilon_{AB}= 1.5 \epsilon_{AA}$, $\epsilon_{BB}=0.5 \epsilon_{AA}$ with a number density of  $\rho=1.2$ in accordance with the original passive Kob-Andersen model~\cite{kob95,bruning08}. The Lennard-Jones potential was truncated at $r^{c}_{\alpha \beta}=2.5 \sigma_{\alpha \beta}$ with a constant and a quadratic term to make the potential and its first derivative ({\textit{i.e.}} force) continuous at the cutoff.

\begin{figure}
\includegraphics[height = 0.7\linewidth]{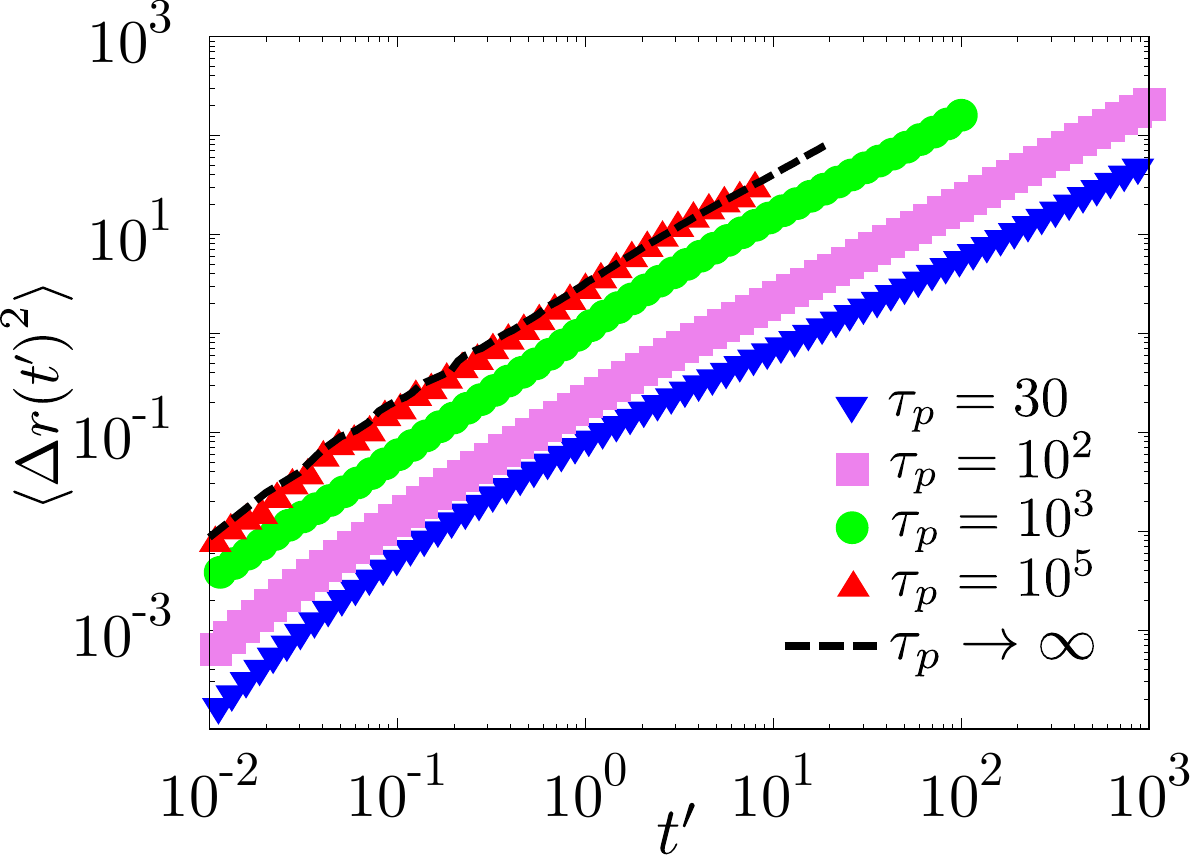}
\caption{Analogue of Fig.~\ref{fig:qt} for the mean squared displacement as a function of scaled time. Convergence to the ADD predictions is again observed for large $\tau_p$.}
\label{fig:msd}
\end{figure}

\begin{figure*}
\includegraphics[height =.55\columnwidth]{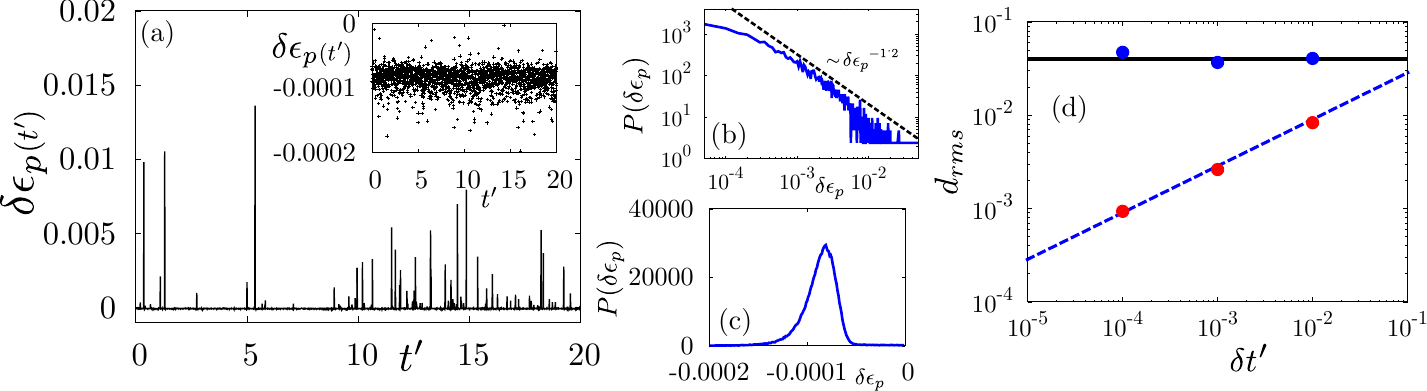}
\caption{(a) Time series of potential energy drops ($\delta \epsilon_p$) at each step of activity driven dynamics, showing both background fluctuation ($\delta \epsilon_p <0$;  see inset for a zoomed in version) and spikes ($\delta \epsilon_p>0$) at the irreversible plastic events. (b) Distribution of positive potential energy drops shows power law decay with an exponent $\sim 1.2$. (c) Distribution of negative potential energy drops shows a narrow uni-modal distribution; these energy ``drops'' indicate small changes of the particle configuration within a smoothly evolving local minimum of $V^{\rm eff}$. (d) Root mean squared particle displacement $d_{rms}$ for plastic events (where $\delta \epsilon_p > 0$, blue points) and for background fluctuations ($\delta\epsilon_p<0$, red points). Lines represent the expected $\delta t'$-independence (black solid) and scaling with $\sqrt{\delta t^{\prime}}$ (blue dashed). Data in sub figures (a,b) taken from Supplementary Information of~\cite{mandalprl20}. }
\label{fig:pot_drop}
\end{figure*}

\section{Numerical Implementation of ADD and Convergence}

To implement ADD we iterate a sequence of two steps in turn: (a) angular update (Eq.(\ref{scaled3})) and (b) minimisation step (Eq.(\ref{minimize})). We first (step (a)) update the propulsion direction for each particle while keeping the position coordinates fixed, using the discretized version of Eq.(\ref{scaled3})
\begin{equation}
    \theta_i(t'+ \delta t')=\theta_i(t') + \sqrt{2 \delta t'}\, \tilde{\eta}_i
    \label{theta_update}
\end{equation}
where $\tilde{\eta}_i$ is a Gaussian random variable with zero mean and unit variance. After each such change in the orientational degrees of freedom we relax the particle configuration to the nearest local energy minimum (of the tilted landscape, step (b)). The total force on each particle vanishes there as prescribed by Eq.(\ref{minimize}). In our numerical implementation, this energy minimization step follows the original inertial dynamics until the root mean square of the total force on each particle, $[\sum_i {\bf f}_i^2/N]^{1/2}$, falls below a small threshold value $F_c$ that we use to decide whether the system has effectively reached a local energy minimum. 

For the minimisation step (b) we use following integration scheme to update the position ${\bf{r}}_i$ of the $i$-th particle in a time step $\Delta t$:
\begin{equation}
    {\bf{r}}_i(t+ \Delta t)=  {\bf{r}}_i(t)+ c_1 {{\bf{v}}_i}(t) + c_2 [{{\bf{f}}_i}(t)+f{\bf n}_i] 
\end{equation}
where $c_1=\frac{m}{\gamma}\left(  1- \Gamma \right)$, $c_2=\frac{m}{\gamma^2}\left( \frac{\gamma \Delta t}{m}- 1 +\Gamma \right)$ and $\Gamma=\exp{(-\frac{\gamma \Delta t}{m})}$. 
For the velocity ${\bf{v}}_i$ of the $i$-th particle the update scheme we use is
\begin{equation}
    {\bf{v}}_i(t+ \Delta t)=  \Gamma {\bf{v}}_i(t)+ \frac{1}{\gamma}\left(1 -\Gamma \right)  [{{\bf{f}}_i}(t)+f{\bf n}_i].
\end{equation}
We continue to update the system using the above steps until the root mean square force threshold is reached, up to some maximum time $t_{\rm step}=250$ in LJ units (defined by setting $\epsilon_{AA}$ and $\sigma_{AA}$ to unity). Because of the presence of the force threshold, the actual time $t_{\rm min}<t_{\rm step}$ taken for the minimization is not fixed but depends on the positions and velocities of the particles at the beginning of the minimization as well as the orientations of active forces; recall that the latter are fixed during the minimization dynamics. The computational advantage of ADD discussed above can now be stated more explicitly by saying that, for large $\tau_p$,  the minimization times $t_{\rm min}$ are significantly smaller than the (unscaled) time interval $\delta t' \tau_p$ corresponding to the update of the propulsion forces in step (a), see Eq.(\ref{theta_update}); in fact in the limit $\tau_p\to\infty$ one has $t_{\rm min}/(\delta t'\tau_p)\to 0$. Fig.~\ref{fig:ke} shows the different time intervals graphically; $t_{\rm min}$ for one ADD step is indicated by the double arrows.

The dynamics discussed above become an exact implementation of ADD in the joint limit of unlimited precision in the  minimization step (b) (in other words a zero threshold $F_c$ on the root mean square force) and vanishing time step for the update of the propulsion force directions, {\textit{i.e.}}\ $\delta t' \to 0$. To check convergence to these limits, 
we first choose an appropriate $\delta t'$ (at fixed force threshold). We do this by running a set of simulations with decreasing $\delta t'$ until we observe convergence of the two-point correlation function $Q$ defined in the next section; this occurs for $\delta t' \le 10^{-2}$. In order to determine the force threshold $F_c$ we similarly check for fixed $\delta t'=10^{-2}$ that the two-point correlation function $Q$ becomes independent of $F_c$ for $F_c \le 10^{-7}$. Based on these observations we use the parameter values $\delta t'=10^{-2}$ and $F_c=10^{-8}$ for all ADD simulations, including a safety margin of one order of magnitude for $F_c$ as this has very little effect on overall computation time.

\section{Comparison with direct simulations}

We benchmark ADD against direct simulations with large $\tau_p$ in terms of both a two-point correlation function $Q(t)$ and the particles' mean squared displacement. The definition of $Q(t)$ is
\begin{equation}
Q(t) =\frac{1}{N} \left\langle  \sum_{i} q(\mid{\bf r}_i(t)- {\bf r}_i(0)\mid)\right\rangle
\end{equation}
where
\begin{equation}
q(x)=
\left\{
        \begin{array}{ll}
                1  & \mbox{if } x \leq b\\
                0  & \mbox{otherwise}
        \end{array}
\right.
\end{equation}
and we choose $b=0.3$ (in units of $\sigma_{AA}$). The mean squared displacement (MSD) is defined as $\langle {\Delta r(t)}^2\rangle=\frac{1}{N} \left\langle  \sum_{i} (\mid{\bf r}_i(t)- {\bf r}_i(0)\mid)^2 \right\rangle$ where as before $\mathbf{r}_i(t)$ is the position of particle $i$ at time $t$. We compare the data for both $Q(t)$ and $\langle {\Delta r(t)}^2\rangle$ from standard simulations~\cite{mandal20} with different $\tau_p$ and plot them as a function of $t'=t/\tau_p$. The trend one observes (see Fig.~\ref{fig:qt} and Fig.~\ref{fig:msd}) indicates that these two point quantities converge in the large persistence time limit. More importantly for our purposes, the limiting behaviours for large $\tau_p$ of both $Q(t)$ and $\langle {\Delta r(t)}^2\rangle$ are entirely consistent with the results predicted by ADD. This confirms that the dynamics obtained from ADD correctly captures  the asymptotic limit of $\tau_p \to \infty$ at fixed $t'=t/\tau_p$ while being significantly faster, in our concrete case by a factor of about an order of magnitude compared to standard simulations at $\tau_p=10^4$.

\section{Events during Activity Driven Dynamics}

Having established ADD as the correct description of the large $\tau_p$-dynamics of active glasses both theoretically and by numerical benchmarking, we next use the method to study the statistics of plastic events in the steady state. As explained above, an advantage of ADD is that it gives us a clean separation between smooth parts of the dynamics, where the particle configuration tracks a gradually evolving local minimum of the potential energy tilted by active forces (see Eq.(\ref{Veff_def})), and instantaneous plastic events where the existing local minimum becomes unstable and the particle configuration rearranges irreversibly to relax to a new minimum. The sketch in Fig.~\ref{fig:tilting} illustrates the distinction using a simple two dimensional energy landscape schematic: the varying tilt of $V^{\rm eff}$ either keeps the particle configuration close to the previous local minimum in any time step or it takes the system away from this original minimum to a new one, signifying an irreversible plastic rearrangement.

To detect for any given time step of duration $\delta t'$ in ADD whether a plastic event has occurred, we calculate the reduction in the tilted potential energy (see Eq.(\ref{Veff_def})) with the active force directions fixed to their values at the {\em beginning} of the time step. We call this quantity (per particle) the energy drop $\delta\epsilon_p$. To aid in the analysis we also measure the root mean squared displacement $d_{\rm rms}$ of particles with $\delta t'$. In a time step where a stable local minimum is changing smoothly, $\delta\epsilon_p$ must be {\em negative}: the particle configuration at the beginning of the time step is at a local minimum of $V^{\rm eff}$ by construction, and as we are keeping the shape of the tilted potential fixed in the definition of $\delta\epsilon_p$, the new configuration at the end of the time step must have a higher $V^{\rm eff}$ and hence $\delta\epsilon_p<0$. The example results from a time series of $\delta\epsilon_p$  confirm this (see Fig.~\ref{fig:pot_drop}a,c). In such a smooth time step one also expects that the particle displacements scale linearly with the changes in the propulsion force directions, so that $d_{\rm rms} \sim  {\delta t'}^{1/2}$. This is what we see in the ADD simulations (Fig.~\ref{fig:pot_drop}d). As the tilted potential energy increases quadratically from a local minimum, we find from $d_{\rm rms}\sim {\delta t'}^{1/2}$ the further estimate $-\delta\epsilon_p \sim d_{\rm rms}^2 \sim \delta t'$, which we also find to be confirmed (data not shown).
\begin{figure}
\includegraphics[height =1.3\columnwidth]{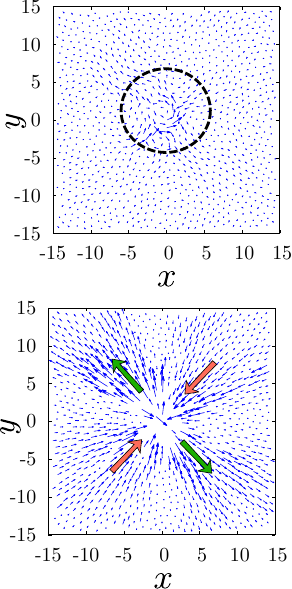}
\caption{(Top) Spatial map of displacement vectors (scaled by a factor of $2$) for an Eshelby-like event, shown circled by a black dashed line. (Bottom) The displacement field scaled by a factor of $20$ for better visualisation, with displacements for particles at the core removed to highlight the far field behaviour, clearly shows the expected dipolar structure.}
\label{fig:event_Eshelby}
\end{figure}
\begin{figure}
\includegraphics[height =1.3\columnwidth]{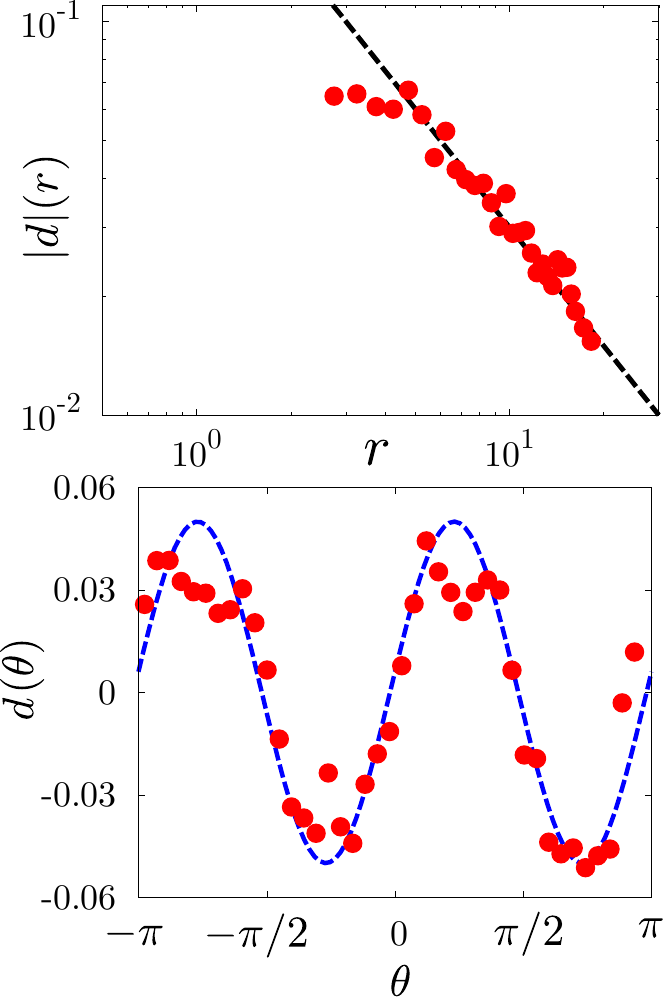}
\caption{(Top) Radial decay of the displacement field in a plastic event, averaged azimuthally, shows $\sim 1/r$ behaviour consistent with the prediction from elasticity theory. (Bottom) The angular dependence of the radial displacements is consistent with dipolar behaviour $\cos(2 \theta)$ (blue dashed line).}
\label{fig:eshelby}
\end{figure}
If on the other hand a plastic event occurs within a time step $\delta t'$, we expect to see values of both $\delta\epsilon_p$ and $d_{\rm rms}$ that do not decrease with $\delta t'$. Also, $\delta\epsilon_p$ will be positive as the system relaxes from a local  minimum that has become unstable to another, lower one. Both expectations are confirmed by our numerical data (see Fig.~\ref{fig:pot_drop}). Note that $d_{\rm rms}$ and $\delta \epsilon_p$ are independent of $\delta t'$ but somewhat smaller than $O(1)$: this makes sense as while we expect maximum particle displacements comparable to the particle diameter in a plastic event, such events are typically localized in space (more on this later) and so only a fraction of particles effectively contributes to $d_{\rm rms}$ and $\delta\epsilon_p$. 

Summarizing, we can identify plastic events in ADD time steps using the energy drop $\delta\epsilon_p$ that we have defined: $\delta\epsilon_p>0$ means that an irreversible particle rearrangement has occurred, while negative values of $\delta\epsilon_p$ indicate smooth, reversible dynamics. Accordingly, in the histogram of $\delta\epsilon_p$ (Fig.~\ref{fig:pot_drop}b,c) we observe a clear peak around negative $\delta\epsilon_p$ that is well separated from the distribution of positive energy drops. As pointed out above, the root mean squared particle displacements in the two types of dynamics -- identified according to the sign of $\delta\epsilon_p$ -- then also scale differently with $\delta t'$ (Fig.~\ref{fig:pot_drop}d).

\begin{figure}
\includegraphics[height =0.6\columnwidth]{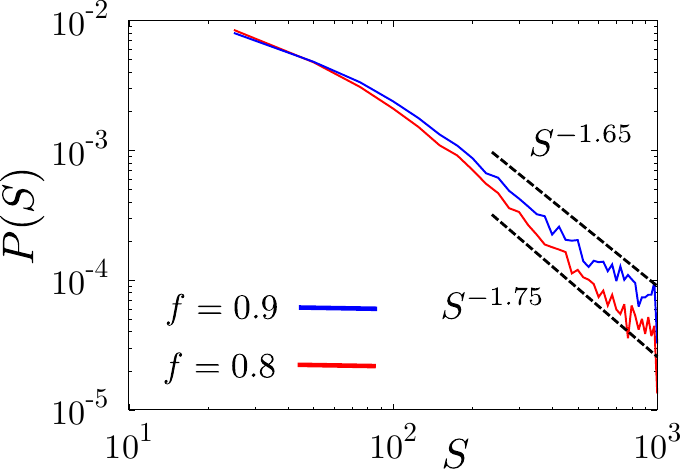}
\caption{Distribution of plastic event sizes for two different active force amplitudes $f=0.8,0.9$, showing a clear power law decay in both cases.}
\label{fig:eventsize}
\end{figure}

We conclude this initial discussion of plastic events by commenting on the 
conceptual similarity between our activity driven dynamics (ADD) and Athermal Quasistatic Shear (AQS)~\cite{maeda78,maeda80,maeda81,maloney04,maloney06} as used to understand the behaviour of glasses under slow shear. Whereas during ADD we make incremental changes in the orientational degrees of freedom determining the active forces, in AQS the analogue is the incremental shear deformation of the system. In both cases an energy minimization follows, which can cause destabilization of local potential energy minima and hence irreversible events. But there is an important distinction between ADD and AQS: in ADD the slow perturbation of the system -- what we have called the tilting of the energy landscape -- is a random process, coming from the diffusive dynamics of the propulsion force orientations. In AQS, on the other hand, the steady shear perturbation has no random component and essentially keeps ``pulling'' the system in the same direction. We have  recently shown that this can lead to very different physical behaviour~\cite{mandalprl20}: at moderate $f$ ADD can facilitate aging while AQS ``interrupts'' ~\cite{kurchan97,berthier01,bonn03} the aging process and instead leads to a stationary state. Even when ADD reaches a stationary state (at higher active force amplitudes $f$), the distribution of (positive) potential energy drops $\delta\epsilon_p$ follows a different power law than observed for AQS~\cite{mandalprl20}. 

We next turn to an analysis of the spatial structure of plastic events in ADD. We will find that again there are similarities here to AQS: the events are typically of Eshelby type, consisting of a core of large plastic displacements with nearly elastic deformations outside the core. An  example by way of orientation is shown in Fig.~\ref{fig:event_Eshelby}.

\section{Statistics of Eshelby-like events}

As before we use ADD to study the steady state dynamics at moderate active force amplitudes $f$. Events are identified as time steps with positive potential energy drops as explained in the previous section. We  determine for each event the core with the largest displacements (see Fig.~\ref{fig:event_Eshelby}) and the orientation of the event, i.e.\ the direction where in the far field the deformation is most strongly extensional. In steady shear as explored by AQS the extensional axis of plastic events tends to be oriented at an angle of $\pi/4$ to the flow direction, as set by the shear geometry. 
In ADD, on the other hand, the random changes of active force directions that cause plastic events have no preferred spatial direction and accordingly we find that the orientations of the Eshelby-like events that occur are distributed uniformly between $0$ and $2 \pi$ (data not shown).

Looking more closely into the Eshelby-like structure, we find a decrease of radial displacements  $d(r)$ with distance $r$ from the core  as $|d|\sim r^{-1}$ (see Fig.~\ref{fig:eshelby}). This matches with the analytical prediction from elasticity theory, which predicts a scaling with $r^{-(1+{d}/{2})}$~\cite{barrat18} for the stress profile; as stress is proportional to displacement {\em gradients} the displacement must then scale with an exponent that is larger by one, {\em i.e.}\ as $r^{1-2}=r^{-1}$ in $d=2$ dimensions. Note that the results in Fig.~\ref{fig:eshelby} relate to a single plastic event; to reduce statistical error we have averaged azimuthally, {\em i.e.}\ over all particles within each radial bin $[r,r+dr]$.

Conversely we have also explored the azimuthal variation of the radial displacement component, now averaging over particles at all distances $r$. Again (see Fig.~\ref{fig:eshelby}) we find a good match with the prediction for Eshelby-like events, with the azimuthal variation being of the form $d(\theta) \sim \cos(2 \theta)$ ~\cite{barrat18}.

\begin{figure}
\includegraphics[height =0.6\columnwidth]{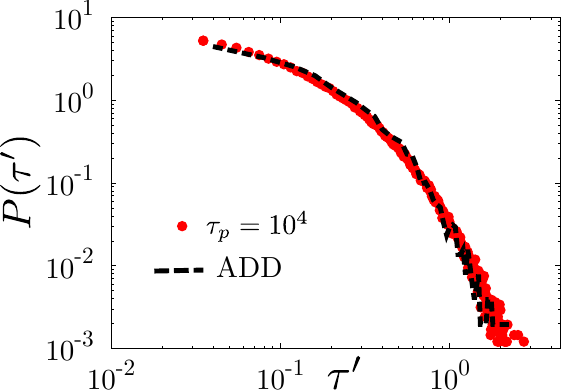}
\caption{Distribution of inter event time scale at small activity $f=0.8$ limit from usual simulation (red points) at $\tau_p=10^4$ and same quantity from activity driven dynamics (black dashed line) shows that ADD can capture the inter event spacing correctly.}
\label{fig:interevent}
\end{figure}

For further insight into the dynamics of plastic events we have analysed the temporal spacing between events using ADD; as the events (defined as before by $\delta \epsilon_p > 0$) are instantaneous in the ADD scheme we can directly measure the time $\tau'$ between any two successive events. The distribution $P(\tau')$ that results (see Fig.~\ref{fig:eventsize}) has the form of a power law with an exponential cut-off~\cite{mandal20}. For comparison we have determined the analogous distribution from a standard  simulation~\cite{mandal20} at $\tau_p=10^4$ and then converted the inter-event times $\tau$ to scaled time $\tau'=\tau/\tau_p$. Fig.~\ref{fig:interevent} shows the comparison between the two approaches; we again find very good overlap, confirming once more the correctness of the ADD method. 

Finally we consider the statistics of event sizes ($S$). We choose two different values of the active force amplitude, $f=0.8, 0.9$, which are close to the boundary between intermittent liquid and dynamical arrest in our system~\cite{mandal20}. We define event size by counting the number of particles that participate in a plastic rearrangement~\cite{srikanth17}. A participating particle is defined in this context as one that moves a distance $\delta r >0.1$ in a single ADD step. The resulting distribution $P(S)$ shows a power law decay (see Fig.~\ref{fig:eventsize}) with an exponent close to the mean field value $-\frac{3}{2}$ ~\cite{barrat18}, at least within the accuracy here and without studying in more detail potential finite size effects. Intriguingly, the observed behaviour exhibits similarities with the event size distribution observed for oscillatory shear simulation across the yielding transition~\cite{srikanth17}, a connection that poses an interesting question for future work.

\section{Discussion}

Active glassy systems in the large persistence time limit $\tau_p \to \infty$ (or extreme active matter systems) exhibit many fascinating dynamical behaviours~\cite{mandal20,mandalprl20}. In this paper we have demonstrated an efficient algorithm to explore this particular limit, which we refer to as activity driven dynamics (ADD). We have discussed in detail the idea behind the simulation scheme and also the details of its implementation. We have also explored briefly the convergence with respect to finite energy minimization accuracy and finite scaled timestep, and have established that ADD can reliably reproduce the dynamics seen in standard simulations for large $\tau_p$, {\em e.g.}\ for mean-squared displacements, two-point correlation functions and the distribution of time intervals between plastic events. In the last two sections we then demonstrated that the plastic events that occur in ADD are of Eshelby type,
showing radial displacements falling off as $\sim r^{-1}$ with distance from the plastic core and varying in dipolar fashion ($\sim \cos(2 \theta)$) in the azimuthal direction. The orientations of the events are distributed isotropically. This is consistent with the absence of any orientational preference in the driving by active force variations and is a key physical difference to driving by quasistatic steady shear. The distribution of event sizes, finally, exhibits a power law scaling close to the transition between intermittent liquid and dynamical arrest, with a quantitative theory for the observed exponent an outstanding question for further research. More broadly, the ADD technique opens the way to systematic exploration of many other properties of extreme active matter in the dense limit, in a manner that avoids computational bottlenecks arising for $\tau_p\to\infty$ in standard simulations.

\paragraph*{Acknowledgement:} We thank Debsankar Banerjee and J\"org Rottler for useful discussions. This project has received funding from the European Union’s Horizon 2020 research and innovation programme under the Marie Skłodowska-Curie grant agreement No 893128. This research was supported in part by the National Science Foundation under Grant No. NSF PHY-1748958.

\bibliography{Activity_Driven_Dynamics}

\begin{thebibliography}{43}%
\makeatletter
\providecommand \@ifxundefined [1]{%
 \@ifx{#1\undefined}
}%
\providecommand \@ifnum [1]{%
 \ifnum #1\expandafter \@firstoftwo
 \else \expandafter \@secondoftwo
 \fi
}%
\providecommand \@ifx [1]{%
 \ifx #1\expandafter \@firstoftwo
 \else \expandafter \@secondoftwo
 \fi
}%
\providecommand \natexlab [1]{#1}%
\providecommand \enquote  [1]{``#1''}%
\providecommand \bibnamefont  [1]{#1}%
\providecommand \bibfnamefont [1]{#1}%
\providecommand \citenamefont [1]{#1}%
\providecommand \href@noop [0]{\@secondoftwo}%
\providecommand \href [0]{\begingroup \@sanitize@url \@href}%
\providecommand \@href[1]{\@@startlink{#1}\@@href}%
\providecommand \@@href[1]{\endgroup#1\@@endlink}%
\providecommand \@sanitize@url [0]{\catcode `\\12\catcode `\$12\catcode
  `\&12\catcode `\#12\catcode `\^12\catcode `\_12\catcode `\%12\relax}%
\providecommand \@@startlink[1]{}%
\providecommand \@@endlink[0]{}%
\providecommand \url  [0]{\begingroup\@sanitize@url \@url }%
\providecommand \@url [1]{\endgroup\@href {#1}{\urlprefix }}%
\providecommand \urlprefix  [0]{URL }%
\providecommand \Eprint [0]{\href }%
\providecommand \doibase [0]{https://doi.org/}%
\providecommand \selectlanguage [0]{\@gobble}%
\providecommand \bibinfo  [0]{\@secondoftwo}%
\providecommand \bibfield  [0]{\@secondoftwo}%
\providecommand \translation [1]{[#1]}%
\providecommand \BibitemOpen [0]{}%
\providecommand \bibitemStop [0]{}%
\providecommand \bibitemNoStop [0]{.\EOS\space}%
\providecommand \EOS [0]{\spacefactor3000\relax}%
\providecommand \BibitemShut  [1]{\csname bibitem#1\endcsname}%
\let\auto@bib@innerbib\@empty
\bibitem [{\citenamefont {Anderson}(1995)}]{anderson95}%
  \BibitemOpen
  \bibfield  {author} {\bibinfo {author} {\bibfnamefont {P.~W.}\ \bibnamefont
  {Anderson}},\ }\bibfield  {title} {\bibinfo {title} {Through the glass
  lightly},\ }\href {https://doi.org/10.1126/science.267.5204.1615-e}
  {\bibfield  {journal} {\bibinfo  {journal} {Science}\ }\textbf {\bibinfo
  {volume} {267}},\ \bibinfo {pages} {1615} (\bibinfo {year}
  {1995})}\BibitemShut {NoStop}%
\bibitem [{\citenamefont {Berthier}\ and\ \citenamefont
  {Biroli}(2011)}]{berthierrmp11}%
  \BibitemOpen
  \bibfield  {author} {\bibinfo {author} {\bibfnamefont {L.}~\bibnamefont
  {Berthier}}\ and\ \bibinfo {author} {\bibfnamefont {G.}~\bibnamefont
  {Biroli}},\ }\bibfield  {title} {\bibinfo {title} {Theoretical perspective on
  the glass transition and amorphous materials},\ }\href
  {https://doi.org/10.1103/RevModPhys.83.587} {\bibfield  {journal} {\bibinfo
  {journal} {Rev. Mod. Phys.}\ }\textbf {\bibinfo {volume} {83}},\ \bibinfo
  {pages} {587} (\bibinfo {year} {2011})}\BibitemShut {NoStop}%
\bibitem [{\citenamefont {Das}(2004)}]{spdas04}%
  \BibitemOpen
  \bibfield  {author} {\bibinfo {author} {\bibfnamefont {S.~P.}\ \bibnamefont
  {Das}},\ }\bibfield  {title} {\bibinfo {title} {Mode-coupling theory and the
  glass transition in supercooled liquids},\ }\href
  {https://doi.org/10.1103/RevModPhys.76.785} {\bibfield  {journal} {\bibinfo
  {journal} {Rev. Mod. Phys.}\ }\textbf {\bibinfo {volume} {76}},\ \bibinfo
  {pages} {785} (\bibinfo {year} {2004})}\BibitemShut {NoStop}%
\bibitem [{\citenamefont {Kirkpatrick}\ and\ \citenamefont
  {Thirumalai}(2015)}]{thirumalai15}%
  \BibitemOpen
  \bibfield  {author} {\bibinfo {author} {\bibfnamefont {T.~R.}\ \bibnamefont
  {Kirkpatrick}}\ and\ \bibinfo {author} {\bibfnamefont {D.}~\bibnamefont
  {Thirumalai}},\ }\bibfield  {title} {\bibinfo {title} {Colloquium: Random
  first order transition theory concepts in biology and physics},\ }\href
  {https://doi.org/10.1103/RevModPhys.87.183} {\bibfield  {journal} {\bibinfo
  {journal} {Rev. Mod. Phys.}\ }\textbf {\bibinfo {volume} {87}},\ \bibinfo
  {pages} {183} (\bibinfo {year} {2015})}\BibitemShut {NoStop}%
\bibitem [{\citenamefont {Dyre}(2006)}]{dyre06}%
  \BibitemOpen
  \bibfield  {author} {\bibinfo {author} {\bibfnamefont {J.~C.}\ \bibnamefont
  {Dyre}},\ }\bibfield  {title} {\bibinfo {title} {Colloquium: The glass
  transition and elastic models of glass-forming liquids},\ }\href
  {https://doi.org/10.1103/RevModPhys.78.953} {\bibfield  {journal} {\bibinfo
  {journal} {Rev. Mod. Phys.}\ }\textbf {\bibinfo {volume} {78}},\ \bibinfo
  {pages} {953} (\bibinfo {year} {2006})}\BibitemShut {NoStop}%
\bibitem [{\citenamefont {Charbonneau}\ \emph {et~al.}(2017)\citenamefont
  {Charbonneau}, \citenamefont {Kurchan}, \citenamefont {Parisi}, \citenamefont
  {Urbani},\ and\ \citenamefont {Zamponi}}]{parisi17}%
  \BibitemOpen
  \bibfield  {author} {\bibinfo {author} {\bibfnamefont {P.}~\bibnamefont
  {Charbonneau}}, \bibinfo {author} {\bibfnamefont {J.}~\bibnamefont
  {Kurchan}}, \bibinfo {author} {\bibfnamefont {G.}~\bibnamefont {Parisi}},
  \bibinfo {author} {\bibfnamefont {P.}~\bibnamefont {Urbani}},\ and\ \bibinfo
  {author} {\bibfnamefont {F.}~\bibnamefont {Zamponi}},\ }\bibfield  {title}
  {\bibinfo {title} {Glass and jamming transitions: From exact results to
  finite-dimensional descriptions},\ }\href
  {https://doi.org/10.1146/annurev-conmatphys-031016-025334} {\bibfield
  {journal} {\bibinfo  {journal} {Annual Review of Condensed Matter Physics}\
  }\textbf {\bibinfo {volume} {8}},\ \bibinfo {pages} {265} (\bibinfo {year}
  {2017})}\BibitemShut {NoStop}%
\bibitem [{\citenamefont {Arceri}\ \emph {et~al.}(2020)\citenamefont {Arceri},
  \citenamefont {Landes}, \citenamefont {Berthier},\ and\ \citenamefont
  {Biroli}}]{berthier20}%
  \BibitemOpen
  \bibfield  {author} {\bibinfo {author} {\bibfnamefont {F.}~\bibnamefont
  {Arceri}}, \bibinfo {author} {\bibfnamefont {F.~P.}\ \bibnamefont {Landes}},
  \bibinfo {author} {\bibfnamefont {L.}~\bibnamefont {Berthier}},\ and\
  \bibinfo {author} {\bibfnamefont {G.}~\bibnamefont {Biroli}},\ }\href@noop {}
  {\bibinfo {title} {Glasses and aging: A statistical mechanics perspective}}
  (\bibinfo {year} {2020}),\ \Eprint {https://arxiv.org/abs/2006.09725}
  {arXiv:2006.09725 [cond-mat.stat-mech]} \BibitemShut {NoStop}%
\bibitem [{\citenamefont {Marchetti}\ \emph {et~al.}(2013)\citenamefont
  {Marchetti}, \citenamefont {Joanny}, \citenamefont {Ramaswamy}, \citenamefont
  {Liverpool}, \citenamefont {Prost}, \citenamefont {Rao},\ and\ \citenamefont
  {Simha}}]{marchetti13}%
  \BibitemOpen
  \bibfield  {author} {\bibinfo {author} {\bibfnamefont {M.~C.}\ \bibnamefont
  {Marchetti}}, \bibinfo {author} {\bibfnamefont {J.~F.}\ \bibnamefont
  {Joanny}}, \bibinfo {author} {\bibfnamefont {S.}~\bibnamefont {Ramaswamy}},
  \bibinfo {author} {\bibfnamefont {T.~B.}\ \bibnamefont {Liverpool}}, \bibinfo
  {author} {\bibfnamefont {J.}~\bibnamefont {Prost}}, \bibinfo {author}
  {\bibfnamefont {M.}~\bibnamefont {Rao}},\ and\ \bibinfo {author}
  {\bibfnamefont {R.~A.}\ \bibnamefont {Simha}},\ }\bibfield  {title} {\bibinfo
  {title} {Hydrodynamics of soft active matter},\ }\href
  {https://doi.org/10.1103/RevModPhys.85.1143} {\bibfield  {journal} {\bibinfo
  {journal} {Rev. Mod. Phys.}\ }\textbf {\bibinfo {volume} {85}},\ \bibinfo
  {pages} {1143} (\bibinfo {year} {2013})}\BibitemShut {NoStop}%
\bibitem [{\citenamefont {Bechinger}\ \emph {et~al.}(2016)\citenamefont
  {Bechinger}, \citenamefont {Di~Leonardo}, \citenamefont {L\"owen},
  \citenamefont {Reichhardt}, \citenamefont {Volpe},\ and\ \citenamefont
  {Volpe}}]{volpe16}%
  \BibitemOpen
  \bibfield  {author} {\bibinfo {author} {\bibfnamefont {C.}~\bibnamefont
  {Bechinger}}, \bibinfo {author} {\bibfnamefont {R.}~\bibnamefont
  {Di~Leonardo}}, \bibinfo {author} {\bibfnamefont {H.}~\bibnamefont
  {L\"owen}}, \bibinfo {author} {\bibfnamefont {C.}~\bibnamefont {Reichhardt}},
  \bibinfo {author} {\bibfnamefont {G.}~\bibnamefont {Volpe}},\ and\ \bibinfo
  {author} {\bibfnamefont {G.}~\bibnamefont {Volpe}},\ }\bibfield  {title}
  {\bibinfo {title} {Active particles in complex and crowded environments},\
  }\href {https://doi.org/10.1103/RevModPhys.88.045006} {\bibfield  {journal}
  {\bibinfo  {journal} {Rev. Mod. Phys.}\ }\textbf {\bibinfo {volume} {88}},\
  \bibinfo {pages} {045006} (\bibinfo {year} {2016})}\BibitemShut {NoStop}%
\bibitem [{\citenamefont {Ramaswamy}(2010)}]{sriram10}%
  \BibitemOpen
  \bibfield  {author} {\bibinfo {author} {\bibfnamefont {S.}~\bibnamefont
  {Ramaswamy}},\ }\bibfield  {title} {\bibinfo {title} {The mechanics and
  statistics of active matter},\ }\href
  {https://doi.org/10.1146/annurev-conmatphys-070909-104101} {\bibfield
  {journal} {\bibinfo  {journal} {Annual Review of Condensed Matter Physics}\
  }\textbf {\bibinfo {volume} {1}},\ \bibinfo {pages} {323} (\bibinfo {year}
  {2010})}\BibitemShut {NoStop}%
\bibitem [{\citenamefont {Klongvessa}\ \emph
  {et~al.}(2019{\natexlab{a}})\citenamefont {Klongvessa}, \citenamefont
  {Ginot}, \citenamefont {Ybert}, \citenamefont {Cottin-Bizonne},\ and\
  \citenamefont {Leocmach}}]{leomach19}%
  \BibitemOpen
  \bibfield  {author} {\bibinfo {author} {\bibfnamefont {N.}~\bibnamefont
  {Klongvessa}}, \bibinfo {author} {\bibfnamefont {F.}~\bibnamefont {Ginot}},
  \bibinfo {author} {\bibfnamefont {C.}~\bibnamefont {Ybert}}, \bibinfo
  {author} {\bibfnamefont {C.}~\bibnamefont {Cottin-Bizonne}},\ and\ \bibinfo
  {author} {\bibfnamefont {M.}~\bibnamefont {Leocmach}},\ }\bibfield  {title}
  {\bibinfo {title} {Active glass: Ergodicity breaking dramatically affects
  response to self-propulsion},\ }\href
  {https://doi.org/10.1103/PhysRevLett.123.248004} {\bibfield  {journal}
  {\bibinfo  {journal} {Phys. Rev. Lett.}\ }\textbf {\bibinfo {volume} {123}},\
  \bibinfo {pages} {248004} (\bibinfo {year} {2019}{\natexlab{a}})}\BibitemShut
  {NoStop}%
\bibitem [{\citenamefont {Klongvessa}\ \emph
  {et~al.}(2019{\natexlab{b}})\citenamefont {Klongvessa}, \citenamefont
  {Ginot}, \citenamefont {Ybert}, \citenamefont {Cottin-Bizonne},\ and\
  \citenamefont {Leocmach}}]{leomach19b}%
  \BibitemOpen
  \bibfield  {author} {\bibinfo {author} {\bibfnamefont {N.}~\bibnamefont
  {Klongvessa}}, \bibinfo {author} {\bibfnamefont {F.}~\bibnamefont {Ginot}},
  \bibinfo {author} {\bibfnamefont {C.}~\bibnamefont {Ybert}}, \bibinfo
  {author} {\bibfnamefont {C.}~\bibnamefont {Cottin-Bizonne}},\ and\ \bibinfo
  {author} {\bibfnamefont {M.}~\bibnamefont {Leocmach}},\ }\bibfield  {title}
  {\bibinfo {title} {Nonmonotonic behavior in dense assemblies of active
  colloids},\ }\href {https://doi.org/10.1103/PhysRevE.100.062603} {\bibfield
  {journal} {\bibinfo  {journal} {Phys. Rev. E}\ }\textbf {\bibinfo {volume}
  {100}},\ \bibinfo {pages} {062603} (\bibinfo {year}
  {2019}{\natexlab{b}})}\BibitemShut {NoStop}%
\bibitem [{\citenamefont {Parry}\ \emph {et~al.}(2014)\citenamefont {Parry},
  \citenamefont {Surovtsev}, \citenamefont {Cabeen}, \citenamefont {O’Hern},
  \citenamefont {Dufresne},\ and\ \citenamefont {Jacobs-Wagner}}]{parry14}%
  \BibitemOpen
  \bibfield  {author} {\bibinfo {author} {\bibfnamefont {B.}~\bibnamefont
  {Parry}}, \bibinfo {author} {\bibfnamefont {I.}~\bibnamefont {Surovtsev}},
  \bibinfo {author} {\bibfnamefont {M.}~\bibnamefont {Cabeen}}, \bibinfo
  {author} {\bibfnamefont {C.}~\bibnamefont {O’Hern}}, \bibinfo {author}
  {\bibfnamefont {E.}~\bibnamefont {Dufresne}},\ and\ \bibinfo {author}
  {\bibfnamefont {C.}~\bibnamefont {Jacobs-Wagner}},\ }\bibfield  {title}
  {\bibinfo {title} {The bacterial cytoplasm has glass-like properties and is
  fluidized by metabolic activity},\ }\href
  {https://doi.org/https://doi.org/10.1016/j.cell.2013.11.028} {\bibfield
  {journal} {\bibinfo  {journal} {Cell}\ }\textbf {\bibinfo {volume} {156}},\
  \bibinfo {pages} {183 } (\bibinfo {year} {2014})}\BibitemShut {NoStop}%
\bibitem [{\citenamefont {Angelini}\ \emph {et~al.}(2011)\citenamefont
  {Angelini}, \citenamefont {Hannezo}, \citenamefont {Trepat}, \citenamefont
  {Marquez}, \citenamefont {Fredberg},\ and\ \citenamefont
  {Weitz}}]{angelini11}%
  \BibitemOpen
  \bibfield  {author} {\bibinfo {author} {\bibfnamefont {T.~E.}\ \bibnamefont
  {Angelini}}, \bibinfo {author} {\bibfnamefont {E.}~\bibnamefont {Hannezo}},
  \bibinfo {author} {\bibfnamefont {X.}~\bibnamefont {Trepat}}, \bibinfo
  {author} {\bibfnamefont {M.}~\bibnamefont {Marquez}}, \bibinfo {author}
  {\bibfnamefont {J.~J.}\ \bibnamefont {Fredberg}},\ and\ \bibinfo {author}
  {\bibfnamefont {D.~A.}\ \bibnamefont {Weitz}},\ }\bibfield  {title} {\bibinfo
  {title} {Glass-like dynamics of collective cell migration},\ }\href
  {https://doi.org/10.1073/pnas.1010059108} {\bibfield  {journal} {\bibinfo
  {journal} {Proceedings of the National Academy of Sciences}\ }\textbf
  {\bibinfo {volume} {108}},\ \bibinfo {pages} {4714} (\bibinfo {year}
  {2011})}\BibitemShut {NoStop}%
\bibitem [{\citenamefont {Janssen}(2019)}]{janssen19}%
  \BibitemOpen
  \bibfield  {author} {\bibinfo {author} {\bibfnamefont {L.~M.~C.}\
  \bibnamefont {Janssen}},\ }\bibfield  {title} {\bibinfo {title} {Active
  glasses},\ }\href {https://doi.org/10.1088/1361-648x/ab3e90} {\bibfield
  {journal} {\bibinfo  {journal} {Journal of Physics: Condensed Matter}\
  }\textbf {\bibinfo {volume} {31}},\ \bibinfo {pages} {503002} (\bibinfo
  {year} {2019})}\BibitemShut {NoStop}%
\bibitem [{\citenamefont {Henkes}\ \emph {et~al.}(2011)\citenamefont {Henkes},
  \citenamefont {Fily},\ and\ \citenamefont {Marchetti}}]{silke11}%
  \BibitemOpen
  \bibfield  {author} {\bibinfo {author} {\bibfnamefont {S.}~\bibnamefont
  {Henkes}}, \bibinfo {author} {\bibfnamefont {Y.}~\bibnamefont {Fily}},\ and\
  \bibinfo {author} {\bibfnamefont {M.~C.}\ \bibnamefont {Marchetti}},\
  }\bibfield  {title} {\bibinfo {title} {Active jamming: Self-propelled soft
  particles at high density},\ }\href
  {https://doi.org/10.1103/PhysRevE.84.040301} {\bibfield  {journal} {\bibinfo
  {journal} {Phys. Rev. E}\ }\textbf {\bibinfo {volume} {84}},\ \bibinfo
  {pages} {040301} (\bibinfo {year} {2011})}\BibitemShut {NoStop}%
\bibitem [{\citenamefont {Ni}\ \emph {et~al.}(2013)\citenamefont {Ni},
  \citenamefont {Stuart},\ and\ \citenamefont {Dijkstra}}]{ni13}%
  \BibitemOpen
  \bibfield  {author} {\bibinfo {author} {\bibfnamefont {R.}~\bibnamefont
  {Ni}}, \bibinfo {author} {\bibfnamefont {M.~A.~C.}\ \bibnamefont {Stuart}},\
  and\ \bibinfo {author} {\bibfnamefont {M.}~\bibnamefont {Dijkstra}},\
  }\bibfield  {title} {\bibinfo {title} {Pushing the glass transition towards
  random close packing using self-propelled hard spheres},\ }\href
  {https://doi.org/10.1038/ncomms3704} {\bibfield  {journal} {\bibinfo
  {journal} {Nature communications}\ }\textbf {\bibinfo {volume} {4}},\
  \bibinfo {pages} {1} (\bibinfo {year} {2013})}\BibitemShut {NoStop}%
\bibitem [{\citenamefont {Berthier}(2014)}]{berthier14}%
  \BibitemOpen
  \bibfield  {author} {\bibinfo {author} {\bibfnamefont {L.}~\bibnamefont
  {Berthier}},\ }\bibfield  {title} {\bibinfo {title} {Nonequilibrium glassy
  dynamics of self-propelled hard disks},\ }\href
  {https://doi.org/10.1103/PhysRevLett.112.220602} {\bibfield  {journal}
  {\bibinfo  {journal} {Phys. Rev. Lett.}\ }\textbf {\bibinfo {volume} {112}},\
  \bibinfo {pages} {220602} (\bibinfo {year} {2014})}\BibitemShut {NoStop}%
\bibitem [{\citenamefont {Berthier}\ and\ \citenamefont
  {Kurchan}(2013)}]{berthier13}%
  \BibitemOpen
  \bibfield  {author} {\bibinfo {author} {\bibfnamefont {L.}~\bibnamefont
  {Berthier}}\ and\ \bibinfo {author} {\bibfnamefont {J.}~\bibnamefont
  {Kurchan}},\ }\bibfield  {title} {\bibinfo {title} {Non-equilibrium glass
  transitions in driven and active matter},\ }\href
  {https://doi.org/10.1038/nphys2592} {\bibfield  {journal} {\bibinfo
  {journal} {Nature Physics}\ }\textbf {\bibinfo {volume} {9}},\ \bibinfo
  {pages} {310} (\bibinfo {year} {2013})}\BibitemShut {NoStop}%
\bibitem [{\citenamefont {Mandal}\ \emph {et~al.}(2016)\citenamefont {Mandal},
  \citenamefont {Bhuyan}, \citenamefont {Rao},\ and\ \citenamefont
  {Dasgupta}}]{mandal16}%
  \BibitemOpen
  \bibfield  {author} {\bibinfo {author} {\bibfnamefont {R.}~\bibnamefont
  {Mandal}}, \bibinfo {author} {\bibfnamefont {P.~J.}\ \bibnamefont {Bhuyan}},
  \bibinfo {author} {\bibfnamefont {M.}~\bibnamefont {Rao}},\ and\ \bibinfo
  {author} {\bibfnamefont {C.}~\bibnamefont {Dasgupta}},\ }\bibfield  {title}
  {\bibinfo {title} {Active fluidization in dense glassy systems},\ }\href
  {https://doi.org/10.1039/C5SM02950C} {\bibfield  {journal} {\bibinfo
  {journal} {Soft Matter}\ }\textbf {\bibinfo {volume} {12}},\ \bibinfo {pages}
  {6268} (\bibinfo {year} {2016})}\BibitemShut {NoStop}%
\bibitem [{\citenamefont {Mandal}\ \emph {et~al.}(2020)\citenamefont {Mandal},
  \citenamefont {Bhuyan}, \citenamefont {Chaudhuri}, \citenamefont {Dasgupta},\
  and\ \citenamefont {Rao}}]{mandal20}%
  \BibitemOpen
  \bibfield  {author} {\bibinfo {author} {\bibfnamefont {R.}~\bibnamefont
  {Mandal}}, \bibinfo {author} {\bibfnamefont {P.~J.}\ \bibnamefont {Bhuyan}},
  \bibinfo {author} {\bibfnamefont {P.}~\bibnamefont {Chaudhuri}}, \bibinfo
  {author} {\bibfnamefont {C.}~\bibnamefont {Dasgupta}},\ and\ \bibinfo
  {author} {\bibfnamefont {M.}~\bibnamefont {Rao}},\ }\bibfield  {title}
  {\bibinfo {title} {Extreme active matter at high densities},\ }\href
  {https://doi.org/10.1038/s41467-020-16130-x} {\bibfield  {journal} {\bibinfo
  {journal} {Nature communications}\ }\textbf {\bibinfo {volume} {11}},\
  \bibinfo {pages} {1} (\bibinfo {year} {2020})}\BibitemShut {NoStop}%
\bibitem [{\citenamefont {Mandal}\ and\ \citenamefont
  {Sollich}(2020)}]{mandalprl20}%
  \BibitemOpen
  \bibfield  {author} {\bibinfo {author} {\bibfnamefont {R.}~\bibnamefont
  {Mandal}}\ and\ \bibinfo {author} {\bibfnamefont {P.}~\bibnamefont
  {Sollich}},\ }\bibfield  {title} {\bibinfo {title} {Multiple types of aging
  in active glasses},\ }\href {https://doi.org/10.1103/PhysRevLett.125.218001}
  {\bibfield  {journal} {\bibinfo  {journal} {Phys. Rev. Lett.}\ }\textbf
  {\bibinfo {volume} {125}},\ \bibinfo {pages} {218001} (\bibinfo {year}
  {2020})}\BibitemShut {NoStop}%
\bibitem [{\citenamefont {Nandi}\ and\ \citenamefont {Gov}(2017)}]{nandi17}%
  \BibitemOpen
  \bibfield  {author} {\bibinfo {author} {\bibfnamefont {S.~K.}\ \bibnamefont
  {Nandi}}\ and\ \bibinfo {author} {\bibfnamefont {N.~S.}\ \bibnamefont
  {Gov}},\ }\bibfield  {title} {\bibinfo {title} {Nonequilibrium mode-coupling
  theory for dense active systems of self-propelled particles},\ }\href
  {https://doi.org/10.1039/C7SM01648D} {\bibfield  {journal} {\bibinfo
  {journal} {Soft Matter}\ }\textbf {\bibinfo {volume} {13}},\ \bibinfo {pages}
  {7609} (\bibinfo {year} {2017})}\BibitemShut {NoStop}%
\bibitem [{\citenamefont {Nandi}\ \emph {et~al.}(2018)\citenamefont {Nandi},
  \citenamefont {Mandal}, \citenamefont {Bhuyan}, \citenamefont {Dasgupta},
  \citenamefont {Rao},\ and\ \citenamefont {Gov}}]{nandi18}%
  \BibitemOpen
  \bibfield  {author} {\bibinfo {author} {\bibfnamefont {S.~K.}\ \bibnamefont
  {Nandi}}, \bibinfo {author} {\bibfnamefont {R.}~\bibnamefont {Mandal}},
  \bibinfo {author} {\bibfnamefont {P.~J.}\ \bibnamefont {Bhuyan}}, \bibinfo
  {author} {\bibfnamefont {C.}~\bibnamefont {Dasgupta}}, \bibinfo {author}
  {\bibfnamefont {M.}~\bibnamefont {Rao}},\ and\ \bibinfo {author}
  {\bibfnamefont {N.~S.}\ \bibnamefont {Gov}},\ }\bibfield  {title} {\bibinfo
  {title} {A random first-order transition theory for an active glass},\ }\href
  {https://doi.org/10.1073/pnas.1721324115} {\bibfield  {journal} {\bibinfo
  {journal} {Proceedings of the National Academy of Sciences}\ }\textbf
  {\bibinfo {volume} {115}},\ \bibinfo {pages} {7688} (\bibinfo {year}
  {2018})}\BibitemShut {NoStop}%
\bibitem [{\citenamefont {Woillez}\ \emph {et~al.}(2020)\citenamefont
  {Woillez}, \citenamefont {Kafri},\ and\ \citenamefont {Gov}}]{nir20}%
  \BibitemOpen
  \bibfield  {author} {\bibinfo {author} {\bibfnamefont {E.}~\bibnamefont
  {Woillez}}, \bibinfo {author} {\bibfnamefont {Y.}~\bibnamefont {Kafri}},\
  and\ \bibinfo {author} {\bibfnamefont {N.~S.}\ \bibnamefont {Gov}},\
  }\bibfield  {title} {\bibinfo {title} {Active trap model},\ }\href
  {https://doi.org/10.1103/PhysRevLett.124.118002} {\bibfield  {journal}
  {\bibinfo  {journal} {Phys. Rev. Lett.}\ }\textbf {\bibinfo {volume} {124}},\
  \bibinfo {pages} {118002} (\bibinfo {year} {2020})}\BibitemShut {NoStop}%
\bibitem [{\citenamefont {Fily}\ and\ \citenamefont
  {Marchetti}(2012)}]{marchetti12}%
  \BibitemOpen
  \bibfield  {author} {\bibinfo {author} {\bibfnamefont {Y.}~\bibnamefont
  {Fily}}\ and\ \bibinfo {author} {\bibfnamefont {M.~C.}\ \bibnamefont
  {Marchetti}},\ }\bibfield  {title} {\bibinfo {title} {Athermal phase
  separation of self-propelled particles with no alignment},\ }\href
  {https://doi.org/10.1103/PhysRevLett.108.235702} {\bibfield  {journal}
  {\bibinfo  {journal} {Phys. Rev. Lett.}\ }\textbf {\bibinfo {volume} {108}},\
  \bibinfo {pages} {235702} (\bibinfo {year} {2012})}\BibitemShut {NoStop}%
\bibitem [{\citenamefont {Takatori}\ and\ \citenamefont
  {Brady}(2015)}]{takatori15}%
  \BibitemOpen
  \bibfield  {author} {\bibinfo {author} {\bibfnamefont {S.~C.}\ \bibnamefont
  {Takatori}}\ and\ \bibinfo {author} {\bibfnamefont {J.~F.}\ \bibnamefont
  {Brady}},\ }\bibfield  {title} {\bibinfo {title} {Towards a thermodynamics of
  active matter},\ }\href {https://link.aps.org/doi/10.1103/PhysRevE.91.032117}
  {\bibfield  {journal} {\bibinfo  {journal} {Phys. Rev. E}\ }\textbf {\bibinfo
  {volume} {91}},\ \bibinfo {pages} {032117} (\bibinfo {year}
  {2015})}\BibitemShut {NoStop}%
\bibitem [{\citenamefont {Levis}\ \emph {et~al.}(2017)\citenamefont {Levis},
  \citenamefont {Codina},\ and\ \citenamefont {Pagonabarraga}}]{levis17}%
  \BibitemOpen
  \bibfield  {author} {\bibinfo {author} {\bibfnamefont {D.}~\bibnamefont
  {Levis}}, \bibinfo {author} {\bibfnamefont {J.}~\bibnamefont {Codina}},\ and\
  \bibinfo {author} {\bibfnamefont {I.}~\bibnamefont {Pagonabarraga}},\
  }\bibfield  {title} {\bibinfo {title} {Active brownian equation of state:
  metastability and phase coexistence},\ }\href
  {https://pubs.rsc.org/en/content/articlepdf/2017/sm/c7sm01504f} {\bibfield
  {journal} {\bibinfo  {journal} {Soft Matter}\ }\textbf {\bibinfo {volume}
  {13}},\ \bibinfo {pages} {8113} (\bibinfo {year} {2017})}\BibitemShut
  {NoStop}%
\bibitem [{\citenamefont {Solon}\ \emph {et~al.}(2018)\citenamefont {Solon},
  \citenamefont {Stenhammar}, \citenamefont {Cates}, \citenamefont {Kafri},\
  and\ \citenamefont {Tailleur}}]{solon18}%
  \BibitemOpen
  \bibfield  {author} {\bibinfo {author} {\bibfnamefont {A.~P.}\ \bibnamefont
  {Solon}}, \bibinfo {author} {\bibfnamefont {J.}~\bibnamefont {Stenhammar}},
  \bibinfo {author} {\bibfnamefont {M.~E.}\ \bibnamefont {Cates}}, \bibinfo
  {author} {\bibfnamefont {Y.}~\bibnamefont {Kafri}},\ and\ \bibinfo {author}
  {\bibfnamefont {J.}~\bibnamefont {Tailleur}},\ }\bibfield  {title} {\bibinfo
  {title} {Generalized thermodynamics of motility-induced phase separation:
  phase equilibria, laplace pressure, and change of ensembles},\ }\href
  {https://doi.org/10.1088/1367-2630/aaccdd} {\bibfield  {journal} {\bibinfo
  {journal} {New Journal of Physics}\ }\textbf {\bibinfo {volume} {20}},\
  \bibinfo {pages} {075001} (\bibinfo {year} {2018})}\BibitemShut {NoStop}%
\bibitem [{\citenamefont {Koumakis}\ \emph {et~al.}(2014)\citenamefont
  {Koumakis}, \citenamefont {Maggi},\ and\ \citenamefont
  {Di~Leonardo}}]{maggi14}%
  \BibitemOpen
  \bibfield  {author} {\bibinfo {author} {\bibfnamefont {N.}~\bibnamefont
  {Koumakis}}, \bibinfo {author} {\bibfnamefont {C.}~\bibnamefont {Maggi}},\
  and\ \bibinfo {author} {\bibfnamefont {R.}~\bibnamefont {Di~Leonardo}},\
  }\bibfield  {title} {\bibinfo {title} {Directed transport of active particles
  over asymmetric energy barriers},\ }\href
  {https://doi.org/10.1039/C4SM00665H} {\bibfield  {journal} {\bibinfo
  {journal} {Soft matter}\ }\textbf {\bibinfo {volume} {10}},\ \bibinfo {pages}
  {5695} (\bibinfo {year} {2014})}\BibitemShut {NoStop}%
\bibitem [{\citenamefont {Marconi}\ and\ \citenamefont
  {Maggi}(2015)}]{marconi15}%
  \BibitemOpen
  \bibfield  {author} {\bibinfo {author} {\bibfnamefont {U.~M.~B.}\
  \bibnamefont {Marconi}}\ and\ \bibinfo {author} {\bibfnamefont
  {C.}~\bibnamefont {Maggi}},\ }\bibfield  {title} {\bibinfo {title} {Towards a
  statistical mechanical theory of active fluids},\ }\href
  {https://doi.org/10.1039/C5SM01718A} {\bibfield  {journal} {\bibinfo
  {journal} {Soft matter}\ }\textbf {\bibinfo {volume} {11}},\ \bibinfo {pages}
  {8768} (\bibinfo {year} {2015})}\BibitemShut {NoStop}%
\bibitem [{\citenamefont {Kob}\ and\ \citenamefont {Andersen}(1995)}]{kob95}%
  \BibitemOpen
  \bibfield  {author} {\bibinfo {author} {\bibfnamefont {W.}~\bibnamefont
  {Kob}}\ and\ \bibinfo {author} {\bibfnamefont {H.~C.}\ \bibnamefont
  {Andersen}},\ }\bibfield  {title} {\bibinfo {title} {Testing mode-coupling
  theory for a supercooled binary lennard-jones mixture i: The van hove
  correlation function},\ }\href {https://doi.org/10.1103/PhysRevE.51.4626}
  {\bibfield  {journal} {\bibinfo  {journal} {Phys. Rev. E}\ }\textbf {\bibinfo
  {volume} {51}},\ \bibinfo {pages} {4626} (\bibinfo {year}
  {1995})}\BibitemShut {NoStop}%
\bibitem [{\citenamefont {Brüning}\ \emph {et~al.}(2008)\citenamefont
  {Brüning}, \citenamefont {St-Onge}, \citenamefont {Patterson},\ and\
  \citenamefont {Kob}}]{bruning08}%
  \BibitemOpen
  \bibfield  {author} {\bibinfo {author} {\bibfnamefont {R.}~\bibnamefont
  {Brüning}}, \bibinfo {author} {\bibfnamefont {D.~A.}\ \bibnamefont
  {St-Onge}}, \bibinfo {author} {\bibfnamefont {S.}~\bibnamefont {Patterson}},\
  and\ \bibinfo {author} {\bibfnamefont {W.}~\bibnamefont {Kob}},\ }\bibfield
  {title} {\bibinfo {title} {Glass transitions in one-, two-, three-, and
  four-dimensional binary lennard-jones systems},\ }\href
  {https://doi.org/10.1088/0953-8984/21/3/035117} {\bibfield  {journal}
  {\bibinfo  {journal} {Journal of Physics: Condensed Matter}\ }\textbf
  {\bibinfo {volume} {21}},\ \bibinfo {pages} {035117} (\bibinfo {year}
  {2008})}\BibitemShut {NoStop}%
\bibitem [{\citenamefont {Maeda}\ and\ \citenamefont
  {Takeuchi}(1978)}]{maeda78}%
  \BibitemOpen
  \bibfield  {author} {\bibinfo {author} {\bibfnamefont {K.}~\bibnamefont
  {Maeda}}\ and\ \bibinfo {author} {\bibfnamefont {S.}~\bibnamefont
  {Takeuchi}},\ }\bibfield  {title} {\bibinfo {title} {Computer simulation of
  deformation in two-dimensional amorphous structures},\ }\href
  {https://doi.org/https://doi.org/10.1002/pssa.2210490233} {\bibfield
  {journal} {\bibinfo  {journal} {Physica Status Solidi (a)}\ }\textbf
  {\bibinfo {volume} {49}},\ \bibinfo {pages} {685} (\bibinfo {year}
  {1978})}\BibitemShut {NoStop}%
\bibitem [{\citenamefont {Kobayashi}\ \emph {et~al.}(1980)\citenamefont
  {Kobayashi}, \citenamefont {Maeda},\ and\ \citenamefont
  {Takeuchi}}]{maeda80}%
  \BibitemOpen
  \bibfield  {author} {\bibinfo {author} {\bibfnamefont {S.}~\bibnamefont
  {Kobayashi}}, \bibinfo {author} {\bibfnamefont {K.}~\bibnamefont {Maeda}},\
  and\ \bibinfo {author} {\bibfnamefont {S.}~\bibnamefont {Takeuchi}},\
  }\bibfield  {title} {\bibinfo {title} {Computer simulation of deformation of
  amorphous cu57zr43},\ }\href {https://doi.org/10.1016/0001-6160(80)90017-6}
  {\bibfield  {journal} {\bibinfo  {journal} {Acta Metallurgica}\ }\textbf
  {\bibinfo {volume} {28}},\ \bibinfo {pages} {1641} (\bibinfo {year}
  {1980})}\BibitemShut {NoStop}%
\bibitem [{\citenamefont {Maeda}\ and\ \citenamefont
  {Takeuchi}(1981)}]{maeda81}%
  \BibitemOpen
  \bibfield  {author} {\bibinfo {author} {\bibfnamefont {K.}~\bibnamefont
  {Maeda}}\ and\ \bibinfo {author} {\bibfnamefont {S.}~\bibnamefont
  {Takeuchi}},\ }\bibfield  {title} {\bibinfo {title} {Atomistic process of
  plastic deformation in a model amorphous metal},\ }\href
  {https://doi.org/10.1080/01418618108236167} {\bibfield  {journal} {\bibinfo
  {journal} {Philosophical Magazine A}\ }\textbf {\bibinfo {volume} {44}},\
  \bibinfo {pages} {643} (\bibinfo {year} {1981})}\BibitemShut {NoStop}%
\bibitem [{\citenamefont {Maloney}\ and\ \citenamefont
  {Lema\^{\i}tre}(2004)}]{maloney04}%
  \BibitemOpen
  \bibfield  {author} {\bibinfo {author} {\bibfnamefont {C.}~\bibnamefont
  {Maloney}}\ and\ \bibinfo {author} {\bibfnamefont {A.}~\bibnamefont
  {Lema\^{\i}tre}},\ }\bibfield  {title} {\bibinfo {title} {Subextensive
  scaling in the athermal, quasistatic limit of amorphous matter in plastic
  shear flow},\ }\href {https://doi.org/10.1103/PhysRevLett.93.016001}
  {\bibfield  {journal} {\bibinfo  {journal} {Phys. Rev. Lett.}\ }\textbf
  {\bibinfo {volume} {93}},\ \bibinfo {pages} {016001} (\bibinfo {year}
  {2004})}\BibitemShut {NoStop}%
\bibitem [{\citenamefont {Maloney}\ and\ \citenamefont
  {Lema\^{\i}tre}(2006)}]{maloney06}%
  \BibitemOpen
  \bibfield  {author} {\bibinfo {author} {\bibfnamefont {C.~E.}\ \bibnamefont
  {Maloney}}\ and\ \bibinfo {author} {\bibfnamefont {A.}~\bibnamefont
  {Lema\^{\i}tre}},\ }\bibfield  {title} {\bibinfo {title} {Amorphous systems
  in athermal, quasistatic shear},\ }\href
  {https://doi.org/10.1103/PhysRevE.74.016118} {\bibfield  {journal} {\bibinfo
  {journal} {Phys. Rev. E}\ }\textbf {\bibinfo {volume} {74}},\ \bibinfo
  {pages} {016118} (\bibinfo {year} {2006})}\BibitemShut {NoStop}%
\bibitem [{\citenamefont {Kurchan}(1997)}]{kurchan97}%
  \BibitemOpen
  \bibfield  {author} {\bibinfo {author} {\bibfnamefont {J.}~\bibnamefont
  {Kurchan}},\ }\bibfield  {title} {\bibinfo {title} {Rheology, and how to stop
  aging},\ }\href@noop {} {\bibfield  {journal} {\bibinfo  {journal} {Jamming
  and Rheology: Constrained Dynamics on Microscopic and Macroscopic Scales}\ ,\
  \bibinfo {pages} {72}} (\bibinfo {year} {1997})}\BibitemShut {NoStop}%
\bibitem [{\citenamefont {Berthier}\ \emph {et~al.}(2001)\citenamefont
  {Berthier}, \citenamefont {Cugliandolo},\ and\ \citenamefont
  {Iguain}}]{berthier01}%
  \BibitemOpen
  \bibfield  {author} {\bibinfo {author} {\bibfnamefont {L.}~\bibnamefont
  {Berthier}}, \bibinfo {author} {\bibfnamefont {L.~F.}\ \bibnamefont
  {Cugliandolo}},\ and\ \bibinfo {author} {\bibfnamefont {J.~L.}\ \bibnamefont
  {Iguain}},\ }\bibfield  {title} {\bibinfo {title} {Glassy systems under
  time-dependent driving forces: Application to slow granular rheology},\
  }\href {https://doi.org/10.1103/PhysRevE.63.051302} {\bibfield  {journal}
  {\bibinfo  {journal} {Phys. Rev. E}\ }\textbf {\bibinfo {volume} {63}},\
  \bibinfo {pages} {051302} (\bibinfo {year} {2001})}\BibitemShut {NoStop}%
\bibitem [{\citenamefont {Abou}\ \emph {et~al.}(2003)\citenamefont {Abou},
  \citenamefont {Bonn},\ and\ \citenamefont {Meunier}}]{bonn03}%
  \BibitemOpen
  \bibfield  {author} {\bibinfo {author} {\bibfnamefont {B.}~\bibnamefont
  {Abou}}, \bibinfo {author} {\bibfnamefont {D.}~\bibnamefont {Bonn}},\ and\
  \bibinfo {author} {\bibfnamefont {J.}~\bibnamefont {Meunier}},\ }\bibfield
  {title} {\bibinfo {title} {Nonlinear rheology of laponite suspensions under
  an external drive},\ }\href {https://doi.org/10.1122/1.1574022} {\bibfield
  {journal} {\bibinfo  {journal} {Journal of Rheology}\ }\textbf {\bibinfo
  {volume} {47}},\ \bibinfo {pages} {979} (\bibinfo {year} {2003})}\BibitemShut
  {NoStop}%
\bibitem [{\citenamefont {Nicolas}\ \emph {et~al.}(2018)\citenamefont
  {Nicolas}, \citenamefont {Ferrero}, \citenamefont {Martens},\ and\
  \citenamefont {Barrat}}]{barrat18}%
  \BibitemOpen
  \bibfield  {author} {\bibinfo {author} {\bibfnamefont {A.}~\bibnamefont
  {Nicolas}}, \bibinfo {author} {\bibfnamefont {E.~E.}\ \bibnamefont
  {Ferrero}}, \bibinfo {author} {\bibfnamefont {K.}~\bibnamefont {Martens}},\
  and\ \bibinfo {author} {\bibfnamefont {J.-L.}\ \bibnamefont {Barrat}},\
  }\bibfield  {title} {\bibinfo {title} {Deformation and flow of amorphous
  solids: Insights from elastoplastic models},\ }\href
  {https://doi.org/10.1103/RevModPhys.90.045006} {\bibfield  {journal}
  {\bibinfo  {journal} {Rev. Mod. Phys.}\ }\textbf {\bibinfo {volume} {90}},\
  \bibinfo {pages} {045006} (\bibinfo {year} {2018})}\BibitemShut {NoStop}%
\bibitem [{\citenamefont {Leishangthem}\ \emph {et~al.}(2017)\citenamefont
  {Leishangthem}, \citenamefont {Parmar},\ and\ \citenamefont
  {Sastry}}]{srikanth17}%
  \BibitemOpen
  \bibfield  {author} {\bibinfo {author} {\bibfnamefont {P.}~\bibnamefont
  {Leishangthem}}, \bibinfo {author} {\bibfnamefont {A.~D.}\ \bibnamefont
  {Parmar}},\ and\ \bibinfo {author} {\bibfnamefont {S.}~\bibnamefont
  {Sastry}},\ }\bibfield  {title} {\bibinfo {title} {The yielding transition in
  amorphous solids under oscillatory shear deformation},\ }\href
  {https://doi.org/10.1038/ncomms14653} {\bibfield  {journal} {\bibinfo
  {journal} {Nature communications}\ }\textbf {\bibinfo {volume} {8}},\
  \bibinfo {pages} {1} (\bibinfo {year} {2017})}\BibitemShut {NoStop}%
\end{thebibliography}%

\end{document}